\documentclass[prb,showpacs,preprintnumbers,preprint,amsmath,amssymb]{revtex4}
\usepackage{graphicx}% Include figure files
\usepackage{dcolumn}% Align table columns on decimal point
\usepackage{bm}% bold math
\usepackage{epstopdf}

\begin{document}

\title{Non-Fermi liquid behavior in a fluctuating valence system, the filled skutterudite compound CeRu$_{4}$As$_{12}$}

\author{R.~E. Baumbach}
\author{P.~C. Ho}
\author{T.~A. Sayles}
\author{M.~B. Maple}

\affiliation{%
Department of Physics and Institute for Pure and Applied Physical
Sciences, University of California at San Diego, La Jolla, CA
92093, USA
}%

\author{R.~Wawryk}
\author{T.~Cichorek}
\author{A.~Pietraszko}
\author{Z.~Henkie}

\affiliation{Institute of Low Temperature and Structure Research,
Polish Academy of Sciences, 50-950 Wroc{\l}aw, Poland}

\date{\today}

\begin{abstract}
Electrical resistivity $\rho$, specific heat C, and magnetic
susceptibility $\chi$ measurements made on the filled skutterudite
CeRu$_4$As$_{12}$ reveal non-Fermi liquid (NFL) T -  dependences
at low T, i.e., $\rho$(T) $\sim$ T$^{1.4}$ and weak power law or
logarithmic divergences in C(T)/T and $\chi$(T). Measurements also
show that the T - dependence of the thermoelectric power S(T)
deviates from that seen in other Ce systems. The NFL behavior
appears to be associated with fluctuations of the Ce valence
between 3$^+$ and 4$^+$ rather than a typical Kondo lattice
scenario that would be appropriate for an integral Ce valence of
3$^+$.
\end{abstract}

\pacs{71.10.Hf, 71.28.+d}
%71.10.Hf Non-Fermi liquid ground states
%71.28.+d Intermediate valence solids : electronic states of

\maketitle

%%%%%%%%%%%%%%%%%%%%%%%%%%%%%%%%%%%%%
\section{Introduction}

The filled skutterudite compounds of the form MT$_4$X$_{12}$ where
M = alkali metal, alkaline earth, lanthanide, actinide, T = Fe,
Ru, Os and X = P, As, Sb have been shown to exhibit a wealth of
strongly correlated electron phenomena including; spin
fluctuations, itinerant ferromagnetism, local moment
ferromagnetism and antiferromagnetism, conventional BCS
superconductivity, unconventional superconductivity, heavy fermion
behavior, and non-Fermi liquid behavior.\cite{Maple03,Aoki05} Many
of these phenomena depend on hybridization between the rare earth
or actinide f - electron states and the conduction electron
states. This trend is evident in the CeT$_4$X$_{12}$ systems, for
which the expected lattice constant value due to the Ln$^{3+}$
lattice contraction at room temperature is strongly depressed for
the phosphides, is not depressed at all for the antimonides, and
is intermediate for the arsenides.\cite{Braun80a,Braun80b} The
affects of hybridization are particularly dramatic for
CeRu$_4$Sb$_{12}$ which, until now, was the only filled
skutterudite known to show non-Fermi liquid (NFL)
behavior.\cite{Takeda00,Bauer01,Dordevic06} In this compound, the
NFL behavior is characterized at low T as follows : resistivity -
$\rho$(T) $\sim$ T$^{1.4}$, specific heat divided by temperature -
C(T)/T $\sim$ -lnT or T$^{-0.58}$, and magnetic susceptibility -
$\chi$(T) $\sim$ T$^{-0.35}$, all of which suggest that
CeRu$_4$Sb$_{12}$ is located near a quantum critical point (QCP).
The value of the lattice constant for CeRu$_4$Sb$_{12}$ at room
temperature indicates that the Ce ions are nearly trivalent,
suggesting a Kondo lattice is the appropriate description of the
physics. However, it should be noted that the unusual evolution of
$\chi$(T) with T signals a more complicated scenario where the Ce
ions may undergo a continuous valence transition from 3$^+$
towards 4$^+$ with decreasing T.

In this paper, we report NFL behavior in the filled skutterudite
CeRu$_4$As$_{12}$, indicating that this compound is also near a
QCP. However, in this case, the lattice constant at room
temperature and the T dependence of $\chi$(T) reveal that the Ce
ions have an intermediate valence between 3$^{+}$ and 4$^{+}$.
Therefore, CeRu$_4$As$_{12}$ is one of only a few compounds where
the appropriate description of the NFL behavior may be a valence
fluctuation rather than a Kondo lattice picture. We present
measurements of $\rho$(T), C(T), $\chi$(T), and thermoelectric
power S(T) for the compound CeRu$_4$As$_{12}$, characterize the
NFL behavior, discuss the evidence for intermediate valence on the
Ce ions, and compare to other Ce based systems where the NFL
behavior may be related to intermediate Ce valence.

\section{Experiment}

Single crystals of CeRu$_4$As$_{12}$ were grown from elements with
purities $\geq$ 99.9 $\%$ by a molten metal flux method at high
temperatures and pressures, the details of which will be reported
elsewhere. \cite{Henkieunpub} After removing the majority of the
flux by distillation, CeRu$_4$As$_{12}$ single crystals of an
isometric form with dimensions up to $\sim$ 0.7 mm were collected
and cleaned in acid in an effort to remove any impurity phases
from the surfaces of the crystals. The crystal structure of
CeRu$_4$As$_{12}$ was determined by X-ray diffraction on a crystal
with dimensions of 0.17 $\times$  0.18 $\times$ 0.23 mm. A total
of 5714 reflections (447 unique, Rint=0.0772) were recorded and
the structure was resolved by the full matrix least squares method
using the SHELX-97 program with a final discrepancy factor
R1=0.0273 [for I $>$ 2$\sigma$(I), wR2=0.0619].\cite{Sheldrick85,
Sheldrick87}

Electrical resistivity $\rho$(T) measurements for 50 mK $<$ T $<$
290 K were performed in a four-wire configuration in zero magnetic
field using a conventional $^4$He cryostat and a $^3$He - $^4$He
dilution refrigerator. Magnetic susceptibility $\chi$(T)
measurements for 1.9 K $<$ T $<$ 300 K were conducted using a
Quantum Design Magnetic Properties Measurement System (MPMS) on a
mosaic of crystals (m $=$ 49 mg) which were mounted on a small
Delrin disc using Duco cement. Specific heat $C(T)$ measurements
for 650~mK $<$ T $<$ 10~K were made using a standard heat pulse
technique on a collection of 11 single crystals (m $=$ $45$~mg)
attached to a sapphire platform with a small amount of Apiezon N
grease in a $^{3}$He semiadiabatic calorimeter. The thermoelectric
power S(T) for 0.5 K $<$ T $<$ 350 K of CeRu$_4$As$_{12}$ single
crystals with length less than  1 mm was determined by the method
described in ref. \cite{Wawryk01}

\section{Results}
Single crystal structural refinement shows that the unit cell of
CeRu$_{4}$As$_{12}$ has the LaFe$_{4}$P$_{12}$-type structure
(\textbf{I}M$\overline{3}$) space group with two formula units per
unit cell, and a lattice constant a = 8.5004(4) \AA, in reasonable
agreement with earlier measurements of a = 8.4908 \AA~ and a =
8.4963 \AA. \cite{Braun80a,Braun80b,Sekine07} Other crystal
structure parameters are summarized in Table I. The displacement
parameter U$_{eq}$ represents the average displacement of an atom
vibrating around its lattice position and is equal to its
mean-square displacement along the Cartesian axes. The
displacement parameters determined for CeRu$_4$As$_{12}$ exhibit
behavior that is typical of the lanthanide filled
skutterudites.\cite{Sales97,Sales99} Table I also indicates that
the Ce and As sites in CeRu$_4$As$_{12}$ may not be fully occupied
since there is $\sim$ 2$\%$ uncertainty in the occupancy factors.
The possible incomplete Ce or As occupancy is even more pronounced
for some other crystals, not shown in table I, where the Ce and As
filling deviates by as much as 3$\%$ from 100$\%$ occupation.

The T - dependence of the electrical resistivity is quite unusual
(fig. 1). The large value of $\rho$ $\sim$ 3.5 m$\Omega$cm at room
temperature suggests that CeRu$_4$As$_{12}$ is a semimetal. Below
300 K, $\rho$(T) decreases monotonically with decreasing T with
negative curvature between 300 K and $\sim$ 150 K, positive
curvature from $\sim$ 150 K to $\sim$ 70 K, negative curvature
between $\sim$ 70 K and $\sim$ 50 K, and a semi-linear region from
$\sim$ 50 K to $\sim$ 10 K. At the lowest temperatures (65 mK $<$
T $<$ 3.5 K), $\rho$(T) can be described by a power law of the
form,
\begin{equation}\label{eqn1}
\rho(T) = \rho_0[1 + AT^n]
\end{equation}
where n = 1.4, A = 0.14 K$^{-n}$, and $\rho$$_0$ $=$ 136
$\mu\Omega$cm for nearly two decades in temperature. The power law
behavior, illustrated in the inset to fig. 1, is consistent with
that seen for other NFL systems.\cite{Stewart01,Maple94,Maple95}
It should also be noted that no superconducting transition was
detected above 65 mK.

Magnetic susceptibility $\chi$(T) data likewise show an unusual T
-  dependence where $\chi$(T) decreases with T down to $\sim$ 90 K
and then increases to low T (fig. 2). The decrease of $\chi$(T)
with decreasing T from $\sim$ 300 K down to $\sim$ 90 K is
consistent with the Ce ions having an intermediate valence between
3$^{+}$ and 4$^{+}$. In this picture, the ratios n$^{3+}$(T) and
n$^{4+}$(T) describe the fraction of Ce ions in each valence
state. As temperature decreases, n$^{4+}$(T) increases subject to
the constraint n$^{3+}$(T) $+$ n$^{4+}$(T) $=$ 1. In the
intermediate valence scenario, the 4f electron shell of each Ce
ion fluctuates between the configuration 4f$^1$ (Ce$^{3+}$) and
4f$^0$ (Ce$^{4+}$) at a frequency $\omega$ $\approx$
k$_B$T$_{vf}$/$\hbar$, where T$_{vf}$ is a characteristic
temperature separating magnetic behavior at high temperatures T
$\gg$ T$_{vf}$ and nonmagnetic behavior at low temperatures T
$\ll$ T$_{vf}$. \cite{Maple71,Wohlleben} For Ce$^{3+}$, there is
one localized 4f-electron and $\chi$(T) should behave as a
Curie-Weiss law modified by the CEF splitting of the ground state
doublet and an excited state quartet by a splitting temperature
T$_{CEF}$ $\sim$ 100 K. For Ce$^{4+}$, there is no localized
f-electron and $\chi$(T) should behave as a Pauli susceptibility.
However, for an intermediate valence state involving a temporal
admixture of the 4f$^1$ and 4f$^0$ configurations, $\chi$(T) is
expected to be intermediate between $\chi$(T) of the Ce$^{3+}$ and
Ce$^{4+}$ integral valence states for T $\gg$ T$_{vf}$ and to
exhibit an enhanced Pauli-like susceptibility for T $\ll$
T$_{vf}$. Magnetic ordering can also occur if the characteristic
temperature for the RKKY interaction is comparable to the valence
fluctuation temperature T$_{vf}$. Thus, down to 90 K, $\chi$(T) in
CeRu$_4$As$_{12}$ is consistent with an intermediate valence
scenario, and one in which the intermediate valence shifts in the
direction of 4$^+$ with decreasing T.

Below 90 K, $\chi$(T) deviates from the typical intermediate
valence scenario by exhibiting an upturn which persists down to
1.9 K. This behavior may indicate the onset of an unusual state
below 90 K, or the presence of paramagnetic impurity ions. The
upturn in $\chi$(T) can be described by various functions at low
temperatures, including a Curie law for 1.9 K $<$ T $<$ 5 K of the
form,

\begin{equation}\label{eqn2}
\chi(T) = \chi_{0} + \frac{C}{T}
\end{equation}
where $\chi$$_0$ $=$ 1.5$\times$10$^{-3}$ cm$^3$/mol and C $=$
4$\times$10$^{-3}$ cm$^3$K/mol. Such a fit is consistent with the
intermediate valence scenario suggested above, where $\chi$$_0$ is
the finite magnetic susceptibility of the intermediate valence
state and the Curie component is due to the small fraction of
remaining Ce$^{3+}$ ions located at defect sites. Another
possibility is that the low temperature Curie-like upturn is due
to other magnetic rare earth impurities such as Gd$^{3+}$. From
the value of the Curie constant obtained above, impurity
concentrations equivalent to $\sim$ 1500 ppm Ce$^{3+}$ or $\sim$
500 ppm Gd$^{3+}$ could account for the Curie contribution to
$\chi$(T). These concentrations seem rather large for paramagnetic
impurities, and therefore suggest that the low temperature
behavior is intrinsic to CeRu$_4$As$_{12}$. As shown below, the T
- dependence of $\chi$(T) below $\sim$ 10 K is consistent with the
NFL behavior observed in $\rho$(T). For 1.9 K $<$ T $<$ 10 K,
$\chi$(T) is described by either a power law or a logarithmic
function (eqs. 3,4) where m $=$ 0.45 and a = 4.6$\times$10$^{-3}$
cm$^3$mol$^{-1}$K$^m$, or b $=$ 4.2$\times$10$^{-3}$ cm$^3$/mol
and c $=$ 1.2$\times$10$^{-3}$ cm$^3$/mol, respectively (fig. 2).
\begin{equation}\label{eqn3}
\chi(T) = aT^{-m}
\end{equation}

\begin{equation}\label{eqn4}
\chi(T) = b - c~ln(T)
\end{equation}

Further analysis of $\chi$(T) can be made by defining the
effective moment $\mu'_{eff}$(T) as shown in fig. 3,
\begin{equation}\label{eqn5}
\mu'_{eff}(T) \equiv \sqrt{\frac{3k_B\chi T}{N_A}}
\end{equation}
Near 2 K, $\mu'_{eff}$ $\sim$ 0.2 $\mu_B$ is severely depressed
from the free ion Hund's rule multiplet value
$\mu_{eff}$(Ce$^{3+}$) $=$ 2.54 $\mu_B$. This result strongly
suggests that at low temperatures the Ce ions are in an
intermediate valence state. The value of $\mu'_{eff}$ increases
steeply up to 50 K, above which it increases more slowly to arrive
at a value of $\mu'_{eff}$(T) $\sim$ 1.8 $\mu_B$ near 300 K, which
is still substantially reduced from the Ce$^{3+}$ Hund's rule
value. Therefore, it appears that throughout the entire
temperature range measured, the Ce ions have a valence
intermediate between 3$^{+}$ and 4$^{+}$

Displayed in fig. 4 are specific heat divided by temperature C/T
vs T$^2$ data. For T $<$ 10 K, the data decrease with temperature
down to 2.6 K, below which there is an upturn that persists to 650
mK. It should be noted that a small feature with a maximum at 2.4
K is also observed. For temperatures 2.6 K $<$ T $<$ 8 K, the data
can be described by,
\begin{equation}\label{eqn6}
C/T = \gamma + \beta T^2
\end{equation}
where $\gamma$ is the electronic specific heat coefficient and
$\beta$ $\propto$ $\theta$$_D$$^{-3}$ describes the lattice
contribution. Fits of eq. 6 to C(T)/T show that $\gamma$ $\sim$ 26
mJ/molK$^2$ and $\theta$$_D$ $\sim$ 156 K. For 650 mK $<$ T $<$
2.6 K, C/T diverges from the T$^2$ T - dependence and increases
with decreasing temperature. This low temperature divergence in
C/T can be described by either a weak power law or a logarithmic
divergence (eqs. 7,8) where $\varphi$ = 100 mJ/mol K$^{2-l}$ and
$l$ = 0.76 or $\phi$ $=$ 101 mJ/mol K$^2$ and $\zeta$ $=$ 84
mJ/mol K$^2$, respectively. Again, the behavior is consistent with
typical NFL phenomena.

\begin{equation}\label{eqn7}
C/T = \varphi T^{-l}
\end{equation}

\begin{equation}\label{eqn8}
C/T = \phi - \zeta lnT
\end{equation}

Fig. 5 shows the T - dependence of the thermoelectric power S(T)
for two samples of CeRu$_{4}$As$_{12}$. In the high temperature
region, there is a two peak structure with a flat high temperature
peak at $^h$T$_{max}$ $=$ 250 K which reaches $^h$S$_{max}$ $=$ 67
V/K and a sharper peak near $^l$T$_{max}$ $\sim$ 75 K which
reaches a maximum value near $^l$S$_{max}$ $=$ 66  V/K. The
positions of the two peaks are sample independent and weakly
sample dependent, respectively. Similar two peak structures have
been observed for various other cerium systems including the heavy
fermion system Ce$_x$Y$_{1-x}$Cu$_{2.05}$Si$_{2}$, \cite{Ocko05}
where x = 0.3, 0.5. In that case, the low-T peak is ascribed to
electron scattering by the ground state doublet of the Ce$^{3+}$
ion while the high temperature peak is due to electronic
scattering from the entire sextet of CEF levels of the Ce$^{3+}$
ion. A more general description of S(T) for the Ce-based
intermetallic systems is given by a theoretical model that takes
into account the CEF splitting of the Ce$^{3+}$ 4f Hund's rule
ground state multiplet and strong Coulumb repulsion of the 4f
electrons.\cite{Zlatic05} However, it is unclear how an
intermediate valence picture should modify these behaviors.

The inset fig. 5i displays the T - dependence of S(T) below 10 K.
The sample with the lower thermoelectric power exhibits an abrupt
step in S(T) by about $\Delta$S$_{5 K}$ $\sim$ 2 V/K near
T$^{\ast}$ = 5.0 K. The step results in a change of sign from
positive above T$^{\ast}$ to negative below. For the sample with
higher S(T), the jump at T$^{\ast}$ is from one nearly linear S(T)
to another with a slightly lower slope. Additionally, a sign
change at 0.9 K is observed for this sample. The S(T) curves for
these two samples converge at an upturn near T $=$ 0.47 K, as
shown in fig. 5(ii) where the average value of S/T is equal to -
0.6 V/K$^2$. It is noteworthy that T$^{\ast}$ coincides with the
upper limit of the temperature range where the electrical
resistivity is described by a power law with n $=$ 1.4 (Fig. 1).
Behnia et al. \cite{Behnia04} have argued that, in the zero
temperature limit, the thermoelectric power should obey the
relation,

\begin{equation}\label{eqn9}
q = \frac{SN_{A}e}{T\gamma}
\end{equation}
where N$_{A}$ is Avogadro's number, e is the electron charge and
the constant N$_{A}$e = 9.65 $\times$ 10$^4$ Cmol$^{-1}$ is the
Faraday number. The dimensionless quantity q corresponds to the
density of carriers per formula unit for the case of a free
electron gas with an energy independent relaxation time. In our
case, q = -2.2 and differs by sign from all eight cerium compounds
considered in Ref. \cite{Behnia04} This unusual result may also be
indicative of an NFL state at low temperatures.

\section{Discussion}
Taken together, measurements of $\rho$(T), $\chi$(T), C(T)/T, and
S(T) apparently reveal an anomalous NFL state for
CeRu$_4$As$_{12}$ at low temperatures. The NFL behavior is
characterized by sub-quadratic power law behavior in $\rho$(T) and
weak power law or logarithmic divergences in $\chi$(T) and C(T)/T.
The NFL phenomena are also evident in S(T) which deviates from
typical behavior previously observed for other Ce compounds and
shows sample dependence at low temperatures. To roughly quantify
the NFL state, a temperature scale T$_0$ may be derived from the
low temperature fits to $\rho$(T), $\chi$(T) and C(T)/T using the
scaled equations,\cite{Maple71,Maple94,Maple95}
\begin{equation}\label{eqn10}
\rho(T) = \rho_0[1 + (T/T_{0\rho})^n]
\end{equation}

\begin{equation}\label{eqn11}
\chi(T) = -Bln(T/T_{0\chi})
\end{equation}

\begin{equation}\label{eqn12}
C(T)/T = -\lambda R/T_{0K} ln(T/bT_{0K}) + \delta
\end{equation}
which yield the values T$_{0\rho}$ $\sim$ 4 K, T$_{0\chi}$ $\sim$
30 K, and T$_{0K}$ $\sim$ 25 K. For the fit to C(T)/T (eq. 12),
the parameters $\lambda$ $=$ 0.251 and b $=$ 0.41 are taken from
the two channel Kondo model which has proven to be a useful
phenomenological description that yields reasonable values of the
scaling temperature T$_{0K}$ for many NFL systems, although is not
strickly appropriate for the present situation. The unusual
behavior seen in S(T) below 5 K also supports the viewpoint that
T$_0$ is in this temperature range.

A comparison to typical heavy fermion systems may also be made by
computing an effective Wilson-Sommerfield ratio R$_W$ $=$
($\pi$$^2$k$^2$$_B$/3$\mu$$^2$$_{eff}$)($\chi$$_0$/$\gamma$).
Since unambiguous choices for $\chi$$_0$ and $\gamma$ are not
obvious, the values are taken at 1.9 K to probe the low
temperature state. The Hund's rule effective magnetic moment
$\mu$$_{eff}$ = 2.54 $\mu_B$ for Ce$^{3+}$ ions is also used. From
these values, R$_W$ is calculated to be $\sim$ 0.5. This value is
similar to those found in f-electron materials that exhibit Fermi
liquid behavior with heavy quasiparticles. However, the
relationship between $\chi$$_0$ and $\gamma$ in an NFL system
remains unclear.

In addition to NFL behavior, CeRu$_4$As$_{12}$ appears to be
characterized by Ce ions with a valence intermediate between 3$^+$
and 4$^+$. This observation is first suggested by the depressed
room temperature lattice constant, as compared to the expected
trivalent lanthanide contraction (LC) value for the
MRu$_4$As$_{12}$ series where M = La - Pr.\cite{Braun80a,Braun80b}
To emphasize this point, a comparison can be made to the related
compounds MT$_4$Sb$_{12}$ and MT$_4$P$_{12}$ where M = La - Nd and
T = Fe, Ru, and Os. For MT$_4$Sb$_{12}$, there is no deviation
from the expected LC value at room temperature and, therefore, the
Ce ions appear to have a valence near 3$^+$, although this does
not preclude the possibility of an intermediate valence state
below room temperature. This viewpoint is supported by XANES
measurements of CeFe$_4$Sb$_{12}$ and CeOs$_4$Sb$_{12}$ where Ce
is shown to be in the trivalent state at room
temperature.\cite{Grandjean00,Cao} In contrast, MT$_4$P$_{12}$
compounds follow a LC except in the case M = Ce where the lattice
constant is severely depressed from the expected value. In this
case, XANES measurements show that the Ce ions in CeT$_4$P$_{12}$
(T = Fe, Ru) are mainly 3$^+$ with a small 4$^+$ contribution at
room temperature.\cite{Lee99,Xue94} It should be noted that XANES
measurements of Ce compounds may underestimate the 4$^+$ compared
to the 3$^+$ component of the Ce valence.\cite{Bianconi82} Since
the lattice constant suppressions for the CeT$_4$As$_{12}$
compounds are between that for the CeT$_4$Sb$_{12}$ and
CeT$_4$P$_{12}$ systems, it is likely that the valence of their Ce
ions are likewise intermediate.

The intermediate valence picture is further supported by the
unusual T - dependence of $\chi$(T) for CeRu$_4$As$_{12}$ which is
similar to that seen for other intermediate valence compounds such
as SmS, SmB$_6$, or La$_{1-x}$Th$_x$ doped with Ce impurities
where the Ce valence evolves from 3$^+$ to an intermediate value
as a function of Th
concentration.\cite{Maple71,Wohlleben,Huber74a,Huber74b} Moreover,
by comparison to CeRu$_4$As$_{12}$, it appears that
CeRu$_4$Sb$_{12}$ may also develop an intermediate valence state
below room temperature.\cite{Takeda00,Bauer01,Dordevic06} Although
it appears that the Ce ions in CeRu$_4$As$_{12}$ have valence
between 3$^+$ and 4$^+$, additional characterization such as XANES
measurements are required to verify this hypothesis. It should
also be noted that a Kondo volume collapse scenario could, in
principle, account for the reduced lattice constant.\cite{Allen82}

In typical discussions of NFL physics in f-electron materials, the
phenomena are described in terms of interactions between the
itinerant electrons and the magnetic rare earth ions. Examples
include; (1) nearness to various types of magnetic quantum
critical points where a second-order phase transition is
suppressed to 0 K and quantum fluctuations govern physical
properties, \cite{Hertz76,Millis93} (2) Kondo disorder where a
range of Kondo temperatures are allowed including T$_K$ $=$ 0 K,
\cite{Bernal95,Miranda96,Miranda97} (3) the Griffiths phase model,
\cite{Neto98} and (4) the quadrupolar Kondo model.\cite{Cox87}
These pictures have been moderately successful in describing
numerous NFL systems. \cite{Stewart01}

However, for the single crystal CeRu$_4$As$_{12}$ specimens
reported here, Kondo scenarios are unlikely to describe the NFL
behavior since they require the presence of Ce$^{3+}$ ions and, as
argued above, the Ce ions in this system appear to have a
temperature dependant intermediate valence. As such, we suggest
that CeRu$_4$As$_{12}$ lies near a QCP where valence fluctuations
on the Ce ions give rise to the incipient NFL state. The existence
of a QCP due to a quantum phase transition from integral to
intermediate valence has been conjectured in a few other
compounds, most notably CeCu$_2$Si$_2$, CeCu$_2$Ge$_2$,
CePd$_2$Si$_2$, and the alloy
CeCu$_2$(Si$_{1-x}$Ge$_x$)$_2$.\cite{Holmes04,Vargoz98,Raymond00,Yuan03,Onodera02,Onishi01}
The low pressure region of the $\emph{T - P}$ phase diagram for
these compounds is similar to that seen for heavy fermion
superconductors such as CeIn$_3$ and CeMIn$_5$ (M $=$ Co, Rh)
where antiferromagnetic order is suppressed with increasing
pressure towards a QCP, around which a superconducting dome is
observed at low temperatures.\cite{Walker97,Hegger00,Jeffries05}
Higher pressures then tune the Ce ion from integral to
intermediate valence, resulting in a second superconducting dome
near a QCP associated with the valence transition. The region
surrounding the superconducting dome maximum is also accompanied
by NFL behavior. By comparison, an analogous description may be
applicable for CeRu$_4$As$_{12}$.

It is also of note that a recent study shows that $\rho$(T) for
polycrystalline specimens of CeRu$_4$As$_{12}$ conforms to
activated behavior $\rho$ $\sim$ exp($\Delta$/T) where $\Delta$
$=$ 50 K while $\chi$(T) and C(T)/T are similar to those reported
here. \cite{Sekine07} The transition from a weakly insulating
state for polycrystalline samples to semimetallic NFL behavior for
single crystal samples further signals that CeRu$_4$As$_{12}$ is
near a QCP and that its physical properties are tuned by some
parameter, such as Ce filling or structural disorder, that is
affected by sample growth technique or degree of crystallinity. A
similar situation is seen for CeRhSb$_{1-x}$Sn$_x$, where CeRhSb
is a Kondo semiconductor and CeRhSn exhibits NFL
behavior.\cite{Slebarski05} In this system, it is possible to tune
through a Kondo insulator - NFL quantum critical point as a
function of dopant x. This situation may be analogous to the
polycrystalline - single crystal CeRu$_4$As$_{12}$ weak insulator
- NFL evolution where the tuning parameter is defect site or Ce
concentration.
\section{Summary}

It has been shown that measurements of $\rho$(T), $\chi$(T),
C(T)/T, and S(T) for single crystals of CeRu$_4$As$_{12}$ conform
to NFL phenomena at low temperatures. Moreover, it appears that
the Ce ions are in an intermediate valence state between 3$^+$ and
4$^+$, although this line of reasoning needs to be confirmed by
x-ray absorption near - edge spectroscopy or measurements of the
lattice constant to low T. As such, CeRu$_4$As$_{12}$ may be a
member of a small class of quantum critical materials where a
valence transition is central to the unusual behavior. By analogy
to certain ternary Ce based compounds with the ThCr$_2$Si$_2$
structure (e.g., CeCu$_2$Si$_2$) and the CeRhSb$_{1-x}$Sn$_x$
alloys, it appears that CeRu$_4$As$_{12}$ may have a complicated
multi-dimensional phase diagram which includes an integral valence
- intermediate valence phase transition and a metal - insulator
transition. To test this premise, we plan to explore the
electronic state in CeRu$_4$As$_{12}$ as a function of pressure,
magnetic field, and chemical substitution through electrical
resistivity, specific heat, magnetization, and thermal transport
measurements.

\section{Acknowledgements}
This work was supported by the U. S. Department of Energy under
grant no. DE FG02-04ER46105 and the National Science Foundation
under grant no. NSF DMR0335173.

\section{Tables and Figures}

% Table 1
\begin{table}
\caption{Atomic coordinates, displacement parameters, and
occupancy factors for CeRu$_4$As$_{12}$. U$_{eq}$ is defined as
one-third of the trace of the orthogonalized U$_{ij}$ tensor. }
\begin{center}
\begin{tabular}{|c|c|c|c|c|c|}
\hline atom & x & y & z & U$_{eq}$(\AA$^2\times$10$^{-3}$) & Occupancy Factor \\
\hline Ru & 0.25 & 0.25 & 0.25   & 3(1) & 1.00(2) \\
\hline As & 0.1495(1) & 0.3499(1) & 0   & 4(1) & 0.99(2) \\
\hline Ce & 0 & 0 & 0 & 11(1) & 0.99(2)   \\
\hline
\end{tabular}
\end{center}
\end{table}

\clearpage
% Fig.  1
\begin{figure}%[htbp]
       \includegraphics[width=5.0in]{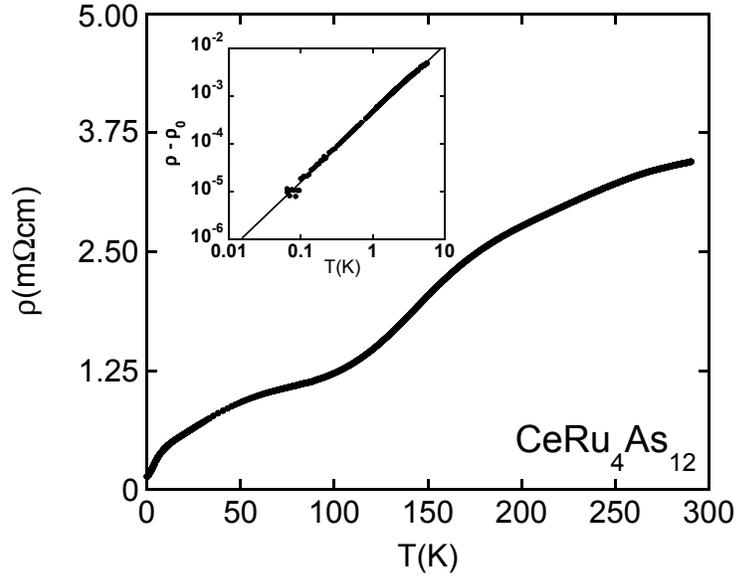}
\caption{Electrical resistivity $\rho$ vs temperature T for
CeRu$_4$As$_{12}$. Inset : log - log plot of $\rho$ - $\rho_0$ vs
T between 65 mK and 10 K, where $\rho_0$ is the residual
resistivity. The data can be described by the expression $\rho$(T)
= $\rho_0$(1 $+$ AT$^n$) with $\rho_0$ = 136 $\mu\Omega$cm, A =
0.14 K$^{-n}$, and n $=$ 1.4 (straight line in the figure).}
\label{Fig1}
\end{figure}

\clearpage
% Fig.  2
\begin{figure}%[htbp]
       \includegraphics[width=5.0in]{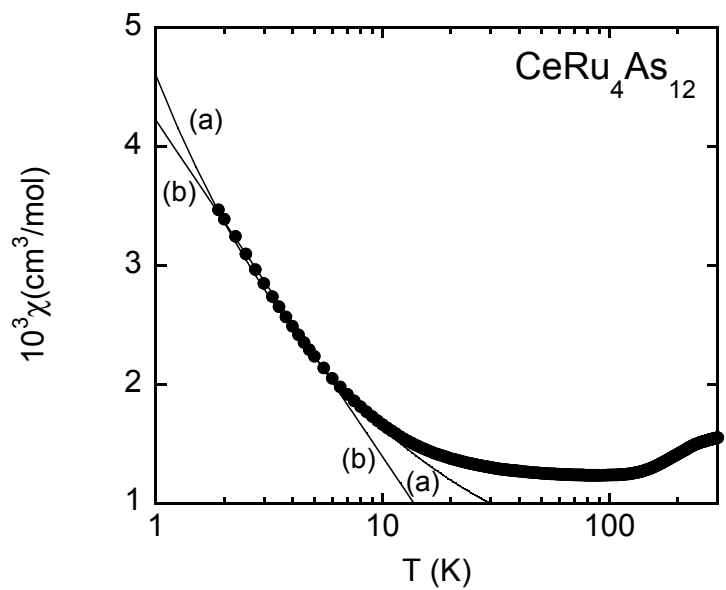}
\caption{Magnetic susceptibility $\chi$ vs temperature T for
CeRu$_4$As$_{12}$ between 2 K and 300 K. For 2 K $<$ T $<$ 10 K,
$\chi$(T) is described by either a power law function (curve a) or
a logarithmic function (curve b) (see text).} \label{Fig2}
\end{figure}

\clearpage
% Fig.  3
\begin{figure}%[htbp]
       \includegraphics[width=5.0in]{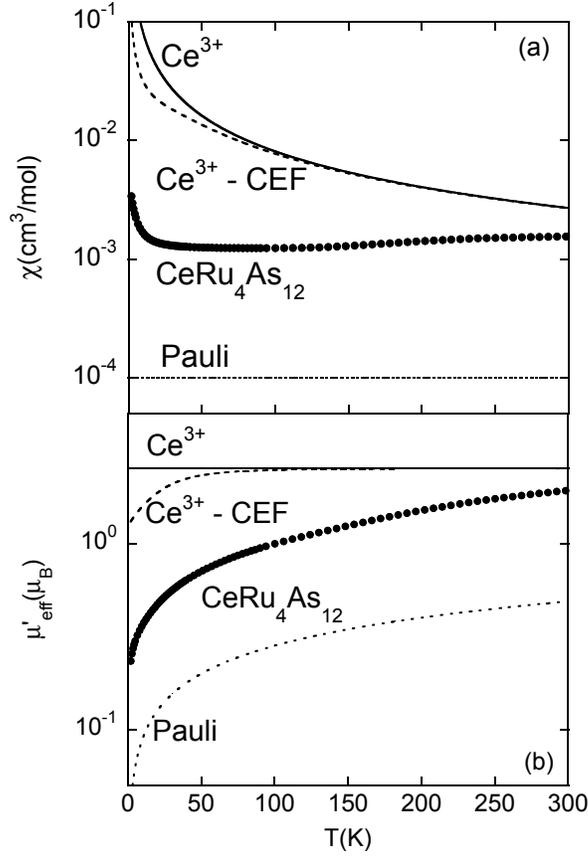}
\caption{(a) Magnetic susceptibility $\chi$ vs temperature T
measured in a magnetic field H = 10 kOe for a CeRu$_4$As$_{12}$
mosaic compared to $\chi$(T) $=$ C/T where C =
N$_A$$\mu$$^2$$_{eff}$/3k$_B$ and $\mu$$_{eff}$ is the Hund's rule
multiplet value of 2.54 $\mu$$_B$ for Ce$^{3+}$ ions (curve
``Ce$^{3+}$"), $\chi$(T) for Ce$^{3+}$ ions for which the
crystalline electric field (CEF) splits the sixfold degenerate
Hund's rule multiplet into a groundstate doublet and an excited
state quartet with a splitting of 100 K (curve ``Ce$^{3+}$ -
CEF"), and a Pauli paramagnetic susceptibility of a typical metal
appropriate for Ce$^{4+}$ ions (curve ``Pauli"). (b) The effective
magnetic moment $\mu$'$_{eff}$ $=$ $\sqrt{\frac{3k_B\chi T}{N_A}}$
for the same cases described in panel a.}\label{Fig3}
\end{figure}

\clearpage
% Fig.  4
\begin{figure}%[htbp]
       \includegraphics[width=5.0in]{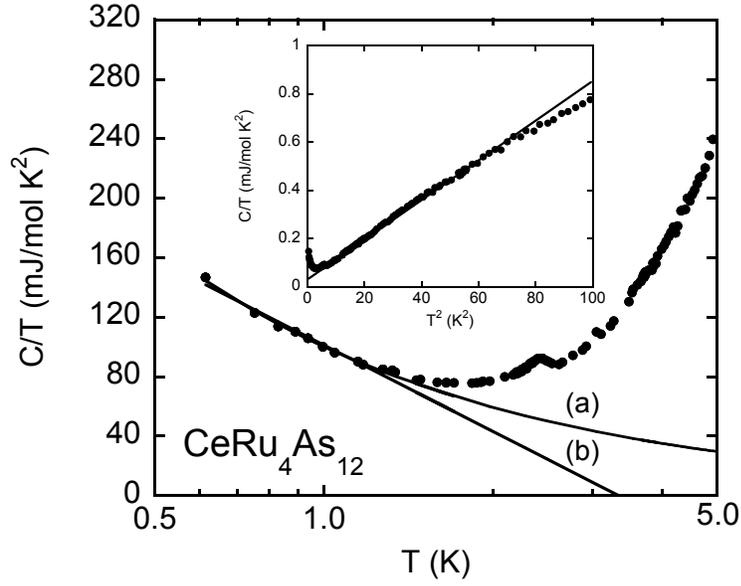}
\caption{Specific heat C divided by temperature T vs T for
CeRu$_4$As$_{12}$. For T $<$ 2.6 K, an upturn is observed which
persists to 650 mK. The divergence conforms to a weak power law
(curve a) or logarithmic function (curve b) (see text). It should
be noted that a small feature is observed with a maximum at 2.4 K.
Shown in the inset is a plot of C/T vs T$^2$ between 0.5 K and 10
K. The straight line is a fit of eq. 6 which yields $\gamma$ = 26
mJ/mol K$^2$ and $\theta$$_D$ = 156 K.} \label{Fig4}
\end{figure}

\clearpage
% Fig.  5
\begin{figure}%[htbp]
       \includegraphics[width=5.0in]{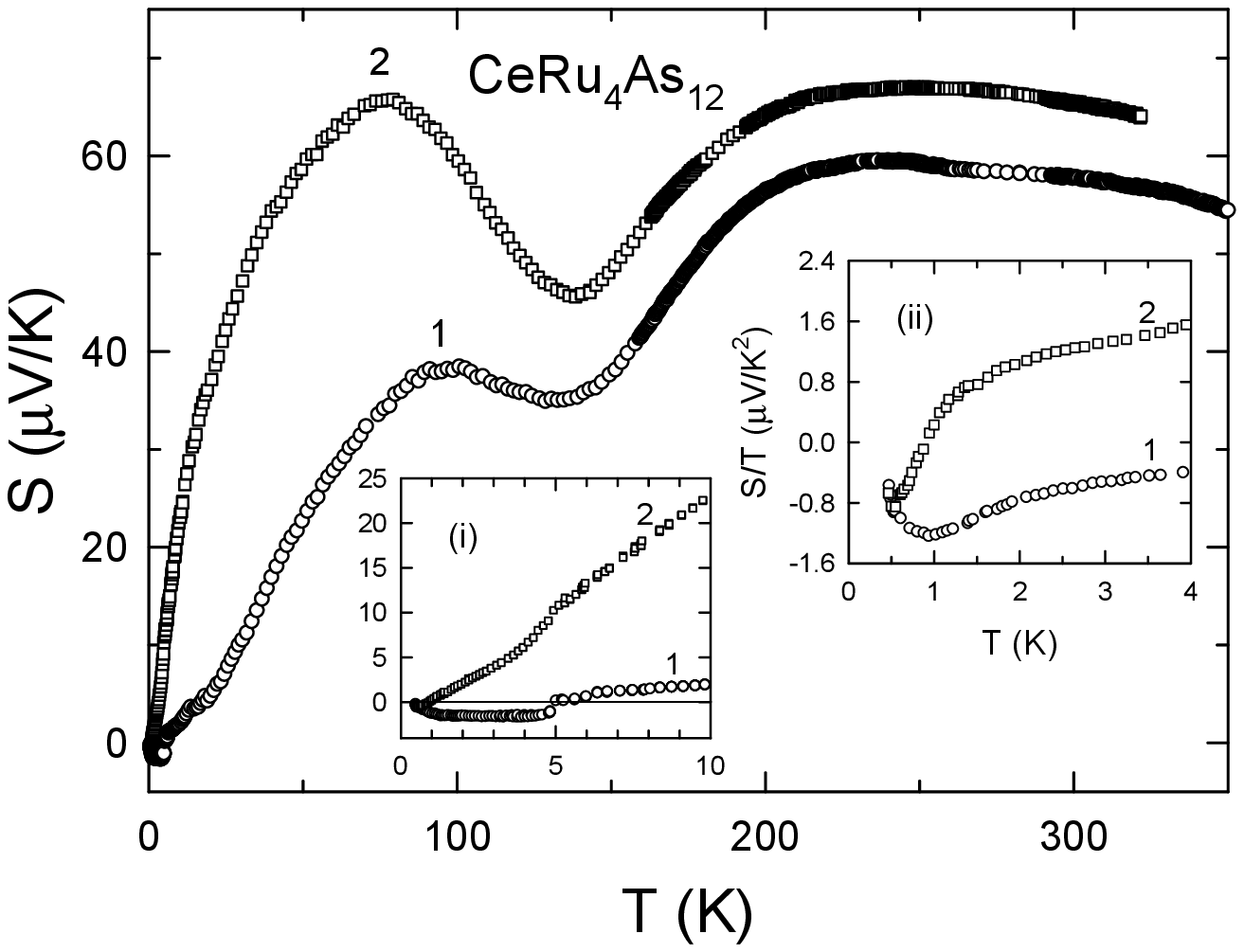}
\caption{Thermoelectric power S vs temperature T for two different
crystals of CeRu$_4$As$_{12}$ from the same batch. Inset (i) shows
S vs T between 0.47 K and 10 K while inset (ii) displays S/T vs T
between 0.47 K and 4 K.} \label{Fig5}
\end{figure}


\begin{thebibliography}{00}

\bibitem{Maple03}
M.~B. Maple, E.~D. Bauer, N.~A. Frederick, P.-C. Ho, W.~M. Yuhasz,
and V.~S. Zapf 2003 \emph{Physica B} {\bf328} $29$.

\bibitem{Aoki05}
Y.~Aoki, H.~Sugawara, H.~Harima, and H.~Sato 2005 \emph{J. Phys.
Soc. Japan} {\bf74} $209$.

\bibitem{Braun80a}
D.~J. Braun and W. Jeitschko 1980 \emph{J. Solid State Chem.}
\textbf{32} 357.

\bibitem{Braun80b} D.~J. Braun and W. Jeitschko 1980 \emph{J. Less
Comm. Met.} \textbf{72} 147.

\bibitem{Takeda00}
M. Takeda and M. Ishikawa 2000 \emph{J. Phys. Soc. Japan} {\bf69}
$868$.

\bibitem{Bauer01}
E.~D. Bauer, A. Slebarski, R.~P. Dickey, E.~J. Freeman, C.
Sirvent, V. Zapf, N.~R. Dilley, and M.~B. Maple 2001 \emph{J.
Phys. Cond. Mat.} \textbf{13} 5183.

\bibitem{Dordevic06}
S.~V. Dordevic, K.~S.~D. Beach, N. Takeda, Y.~J. Wang, M.~B.
Maple, and D.~N. Basov 2006 \emph{Phys. Rev. Lett.} \textbf{96}
017403.

\bibitem{Henkieunpub}
Z. Henkie (unpublished).

\bibitem{Sheldrick85}
G.~M. Sheldrick, in Program for the Solution of Crystal Structures
(University of G\"{o}ttingen, Germany, 1985).

\bibitem{Sheldrick87}
G.~M. Sheldrick, in Program for Crystal Structure Refinement
(University of G\"{o}ttingen, Germany, 1987).

\bibitem{Wawryk01}
R. Wawryk and Z. Henkie 2001 \emph{Phil. Mag. B} \textbf{81} 223.

\bibitem{Sekine07}
C. Sekine, N. Hoshi, K. Takeda, T. Yoshida, I. Shirotani, K.
Natsuhira, M. Wakeshima, and Y. Hinatsu 2007 \emph{J. Magn. Magn.
Matter.} \textbf{310} 260.

\bibitem{Sales97}
B.~C. Sales, D. Mandrus, B.~C. Chakoumakos, V. Keppens, and J.~R.
Thompson 1997 \emph{Phys. Rev. B} \textbf{56} 15081.

\bibitem{Sales99}
B.~C. Sales, B.~C. Chakoumakos, D. Mandrus, J.~W. Sharp, N.~R.
Dilley, and M.~B. Maple 1999 \emph{Mater. Res. Soc. Symp. Proc.}
\textbf{545} 13.

\bibitem{Stewart01}
G.~R. Stewart 2001 \emph{Rev. Mod. Phys.} \textbf{73} 797.

\bibitem{Maple94}
M.~B. Maple, C.~L. Seaman, D.~A. Gajewski, Y. Dalichouch, V.~B.
Barbetta, M.~C. deAndrade, H.~A. Mook, H.~G. Lukefahr, O.~O.
Bernal, and D.~E. MacLaughlin 1994 \emph{J. Low Temp. Phys}
\textbf{95} 225.

\bibitem{Maple95}
M.~B. Maple, M.~C. deAndrade, J. Herrmann, Y. Dalichouch, D.~A.
Gajewski, C.~L. Seaman, R. Chau, R. Movshovich, M.~C. Aronson, and
R. Osborn 1995 \emph{J. Low Temp. Phys} \textbf{99} 224.

\bibitem{Maple71}
M.~B. Maple and D. Wohlleben, \emph{Phys. Rev. Lett.} 1971
\textbf{27} 511.

\bibitem{Wohlleben} M.~B. Maple and D. Wohlleben,
AIP Conference Proceedings (No. 18), Magnetism and Magnetic
Materials - 1073, eds. C. D. Graham, Jr., and J. J. Rhyne, 1974;
p. 447.

\bibitem{Ocko05}
M. Ocko, C. Geibel and F. Steglich 2001 \emph{Phys. Rev. B}
\textbf{64} 195107.

\bibitem{Zlatic05}
V. Zlatic and R. Monnier 2005 \emph{Phys. Rev. B} \textbf{71}
165109.

\bibitem{Behnia04}
K. Behnia, D. Jaccard and J. Flouquet 2004 \emph{J. Phys.:
Condens. Matter} \textbf{16} 5187.

\bibitem{Grandjean00}
F. Grandjean, G.~J. Long,~R. Cortes, D.~T. Morelli, and G.~P.
Meisner 2000 \emph{Phys. Rev. B} \textbf{62} 12569.

\bibitem{Cao}
D. Cao, F. Bridges, R. Baumbach, and M. B. Maple (unpublished).

\bibitem{Lee99}
C.~H. Lee, H. Oyanagi, C. Sekine, I. Shirotani, and M. Ishii 1999
\emph{Phys. Rev. B} \textbf{60} 253.

\bibitem{Xue94}
J.~S Xue, M.~R. Atonio, W.~T. White, L. Soderholm, and S.~M.
Kauzlarich, \emph{J. Alloys and Comp.} 1994 \textbf{207-208} 161.

\bibitem{Bianconi82} A. Bianconi, M. Campagna, and S. Stizza 1982
\emph{Phys. Rev. B} \textbf{25} 2477.

\bibitem{Huber74a} J.~G. Huber, W.~A. Fertig, and M.~B. Maple 1974
\emph{Solid State Comm.} \textbf{15} 453.

\bibitem{Huber74b} J.~G. Huber, J. Brooks, D. Wohlleben, and M.~B.
Maple, \emph{AIP Conference Proceedings (No. 24), Magnetism and
Magnetic Materials}; p. 475 (2 pages) Research Article.

\bibitem{Allen82}
J.~W. Allen and R. Martin 1982 \emph{Phys. Rev. Lett.} \textbf{49}
1106.

\bibitem{Hertz76}
J.~A. Hertz, \emph{Phys. Rev. B} 1973 \textbf{14} 1165.

\bibitem{Millis93}
A.~J. Millis, \emph{Phys. Rev. B} 1993 \textbf{48} 7183.

\bibitem{Bernal95}
O.~O. Bernal, D.~E. MacLaughlin, H.~G. Lukefahr, and B. Andraka
1995 \emph{Phys. Rev. Lett} \textbf{75} 2023.

\bibitem{Miranda96}
E. Miranda, V. Dobrosavljevic, and G. Kotliar 1996 \emph{J. Phys.:
Condens. Matter} \textbf{8} 9871.

\bibitem{Miranda97}
E. Miranda, V. Dobrosavljevic, and G. Kotliar 1997 \emph{Phys.
Rev. Lett.} \textbf{78} 290.

\bibitem{Neto98}
A.~H. Castro Neto, G. Castilla, and B.~A. Jones 1998 \emph{Phys.
Rev. Lett.} \textbf{81} 3531.

\bibitem{Cox87}
D.~L. Cox 1987 \emph{Phys. Rev. Lett.} \textbf{59} 1240.

\bibitem{Yuan03}
H.~Q. Yuan, F.~M. Grosche, M. Deppe, C. Geibel, G. Sparn, and F.
Steglich 2003 \emph{Science} \textbf{302} 2104.

\bibitem{Holmes04}
A. Holmes, D. Jaccard, and K. Miyake 2004 \emph{Phys. Rev. B}
\textbf{69} 024508.

\bibitem{Vargoz98}
E. Vargoz and D. Jaccard 1998 \emph{J. Mag. Mag. Mat.}
\textbf{177-181} 294-295.

\bibitem{Raymond00}
S. Raymond and D. Jaccard 2000 \emph{Phys. Rev. B} \textbf{61}
8679.

\bibitem{Onodera02} A. Onodera, S. Tsuduki, Y. Ohishi, T.
Watanuki, K. Ishida, Y. Kitaoka, and Y. Onuki 2002 \emph{J. Solid
State Comm.} \textbf{123} 113-116.

\bibitem{Onishi01} Y. Onishi and K. Miyake, \emph{J. Phys. Soc.
Japan.} 2001 \textbf{69} 3955.

\bibitem{Walker97}
I.~R. Walker, F.~M. Grosche, D.~M. Freye, and G.~G. Lonzariach,
\emph{Physica C} 1997 \textbf{282-287} 303.

\bibitem{Hegger00}
H. Hegger, C. Petrovic, E.~G. Moshopoulou, M.~F. Hundley, J.~L.
Sarrao, Z. Fisk, and J.~D. Thompson 2000 \emph{Phys. Rev. Lett.}
\textbf{84} 4986.

\bibitem{Jeffries05}
J. Jeffries, N.~A. Frederick, E.~D. Bauer, H. Kimura, V.~S. Zapf,
K.~-D. Hof, and T.~A. Sayles 2005 \emph{Phys. Rev. B} \textbf{72}
024551.

\bibitem{Slebarski05} A. Slebarski and J. Spalek 2005 \emph{Phys. Rev.
Lett.} \textbf{95} 046402.

\end{thebibliography}
\end{document}